\documentclass[ reprint, amsmath,amssymb,prb,aps]{revtex4-2}
\usepackage{color}
\usepackage{verbatim}
\usepackage{lipsum}
\usepackage{hyphenat}
\usepackage{dirtytalk}
\usepackage{graphicx}
\usepackage{dcolumn}
\usepackage{bm}
\usepackage{tabularx}
\usepackage{hyperref}
\usepackage{mathtools}
\usepackage{subfigure}
\usepackage{tabularx}
\usepackage{etoolbox}
\usepackage{colortbl}
\preprint{APS/123-QED}
\newsavebox{\measurebox}
\newcolumntype{l}{X}
\newcolumntype{s}{>{\hsize=0.3\hsize}X}
\newcolumntype{m}{>{\hsize=0.85\hsize}X}

\begin{document} 

\title{Microwave Cavity Mode Optimisation by Background Anti-Resonance Tuning}

\author{Michael T. Hatzon}
\affiliation{Quantum Technologies and Dark Matter Labs, Department of Physics, University of Western Australia, 35 Stirling Highway, Crawley, WA 6009, Australia.}
\email{22873723@student.uwa.edu.au,  michael.tobar@uwa.edu.au}
\author{Eugene N. Ivanov}
\affiliation{Quantum Technologies and Dark Matter Labs, Department of Physics, University of Western Australia, 35 Stirling Highway, Crawley, WA 6009, Australia.}
\author{Jeremy F. Bourhill}
\affiliation{Quantum Technologies and Dark Matter Labs, Department of Physics, University of Western Australia, 35 Stirling Highway, Crawley, WA 6009, Australia.}
\author{Maxim Goryachev}
\affiliation{Quantum Technologies and Dark Matter Labs, Department of Physics, University of Western Australia, 35 Stirling Highway, Crawley, WA 6009, Australia.}
\author{Michael E. Tobar}
\affiliation{Quantum Technologies and Dark Matter Labs, Department of Physics, University of Western Australia, 35 Stirling Highway, Crawley, WA 6009, Australia.}
\email{michael.tobar@uwa.edu.au}

\date{\today}

\begin{abstract}
To derive the best oscillator phase noise when implementing a high-Q resonator, the resonant spectral line-shape must have high contrast and symmetry. Ideally this line-shape is second-order and Lorentzian, however, in a high mode density spectral region, low-Q background spurious modes interact and distort the resonance. For a sapphire-loaded cavity resonator operating with whispering gallery modes confined within the sapphire crystal, we show that this high contrast and symmetry can be achieved by meticulously changing the dimensions of the surrounding metallic cavity shield to tune the background low-Q structures into anti-resonance. This works because the high-Q resonances are primarily defined by the sapphire while the background modes are defined by the cavity shield. Alternatively, it was shown that a similar result can be achieved by exciting the high-Q resonator with a balanced microwave dipole probe in a Mach Zehnder interferometric configuration. The probe was constructed from two separate coaxial electric field probes symmetrically inserted into a cylindrical cavity resonator, from opposite sides with a small gap between them, so they can behave like an active wire dipole antenna. The power into the two separate probes may be matched with an external variable attenuator in one of the arms of the interferometer. Conversely, the phase between the two electric field probes may be changed with an external variable phase shifter, which changes the nature of the field components the probe couples to. The probe couples to the high-Q resonant modes as well as low-Q background modes, which can be made resonant or anti-resonant with respect to the high-Q modes by changing this external phase. When the background modes are in anti-resonance the line shape of the high-Q mode can be made symmetric and with higher contrast. This technique has been applied to both whispering gallery sapphire modes, as well as hollow-cavity resonators, without changing the dimensions of the cavity.

\end{abstract}
\maketitle

\section{Introduction}

High precision low phase noise frequency sources are extremely important in a range of modern applications, from fundamental tests of physics \cite{Nagel:2015dd,Lo16,Goryachev2018,Campbell21,Cat21,Cat23,ivanov2021low}, low-noise measurements \cite{Ivanov:2009aa}, through digital communications, navigation, radar technology to audio processing \cite{Emmerich23}. Amongst the best is the low noise sapphire oscillator both cryogenic, (which implements Pound frequency stabilization) \cite{Woode96,Locke2008,Chaudy20,Fluhr23} and room temperature, (which implements interferometric signal processing for frequency stabilization)\cite{Tobar1995,Ivanov1998,Ivanov2006,Ivanov:2009pv}. Low phase noise resonator-oscillators operate optimally when configured with high-Q resonators with perfectly symmetric line-shape, usually of Lorentzian form, so the phase noise will be filtered optimally outside the bandwidth of the resonator \cite{ivanov2021noise}. It has been shown that any asymmetry, like a Fano type asymmetry \cite{PhysRev.124.1866}, will cause phase noise of an oscillator to be enhanced \cite{ivanov2021noise,ivanov2021low}.

The fact that microwave cavities are multi-mode devices, with probes that couple energy in and out of the cavity, means that high-Q modes exist amongst a background of low-Q modes with significant off-resonance transmission. These low-Q modes also couple to the external probing circuit, often acting to distort the operational modes with a more general Fano type line-shape \cite{tobar1991generalized}. An anti-resonance is defined by a spectral location where the destructive interference between an external driving force and a resonant mode, or multiple modes, results in a local minimum amplitude of oscillation. Previous work has examined an anti-resonance system to probe phase information due to the interference fields of an atom and external pumped fields \cite{PhysRevLett.112.043601}, as well as similar interference effects in microwave-magnon systems \cite{PhysRevB94054403}. Here the effect is purely electromagnetic, showing interference effects between microwave modes similar to that described previously \cite{tobar1991generalized}. 

\begin{figure}[t]
\includegraphics[width=1\columnwidth]{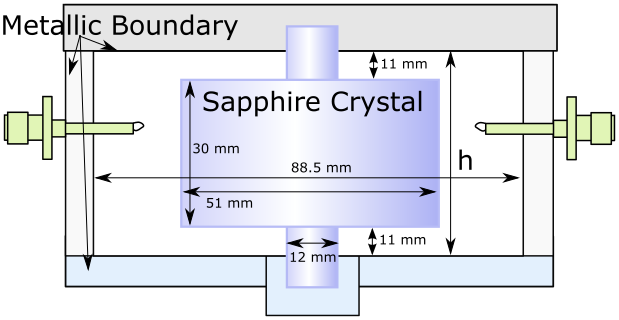}
\caption{The cryogenic sapphire-loaded cavity resonator. In this work the height, $h$, of the cavity was varied, which tuned the background anti-resonances without tuning the sapphire mode. In this work, $\Delta h=0$ corresponded to a value of $h=52.10$ mm.}
\label{EugCavity}
\end{figure}

In this work we investigate ways for an operational high-Q mode to exist in a background anti-resonance. In the first case we examine a high-Q Whispering Gallery (WG) mode sapphire-loaded cavity resonator cooled to near 4 K, as shown in Fig.~\ref{EugCavity}. This type of resonator acts as the frequency determination for the highly stable cooled sapphire oscillator, which may excite an atomic clock at the quantum projection noise limit \cite{Abgrall2016}. For this case we succeeded in realising the operation in a background anti-resonance by modifying the dimensions of the copper cavity that surrounds it, and show that it optimises the contrast and Q-factor of the operational mode. Following this we introduce the concept of a dipole antenna probe with two input arms to excite the cavity resonator. The probing mechanism configuration is similar to a Mach Zehnder interferometer \cite{born2013principles}, as the two arms may have the relative phase and amplitude varied. In this case we show how it can be implemented to improve microwave line shape symmetry and contrast simply through varying the external parameters. In this case we only undertook measurements at room temperature as a proof of concept experiment.

\begin{figure}[h!]
\begin{minipage}{1\columnwidth}
  \includegraphics[width=1\columnwidth]{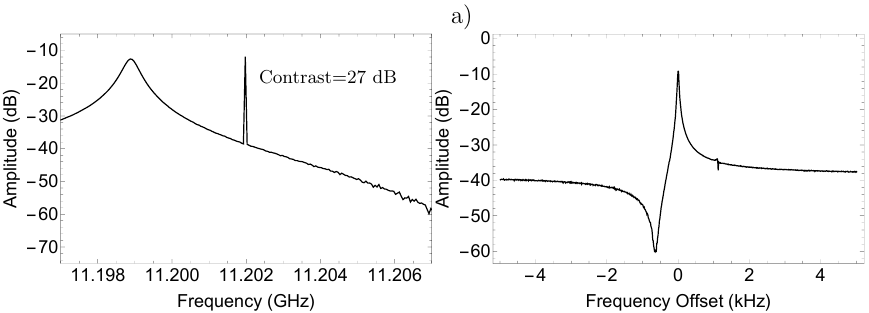}
  \label{Eug3}
\end{minipage}
\begin{minipage}{1\columnwidth}
  \includegraphics[width=1\columnwidth]{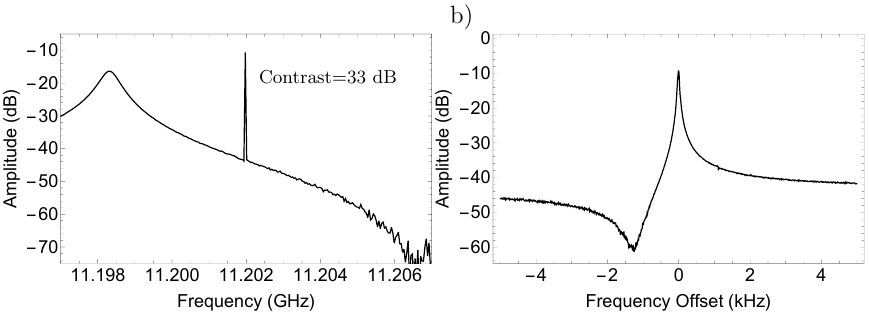}
  \label{Eug4}
\end{minipage}
\begin{minipage}{1\columnwidth}
  \includegraphics[width=1\columnwidth]{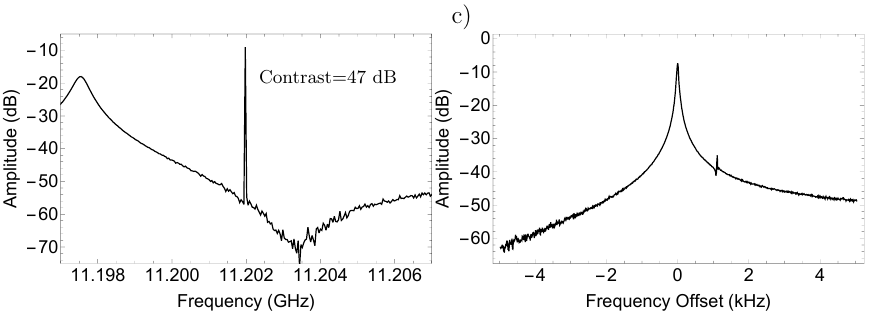}
  \label{Eug5}
\end{minipage}
\begin{minipage}{1\columnwidth}
  \includegraphics[width=1\columnwidth]{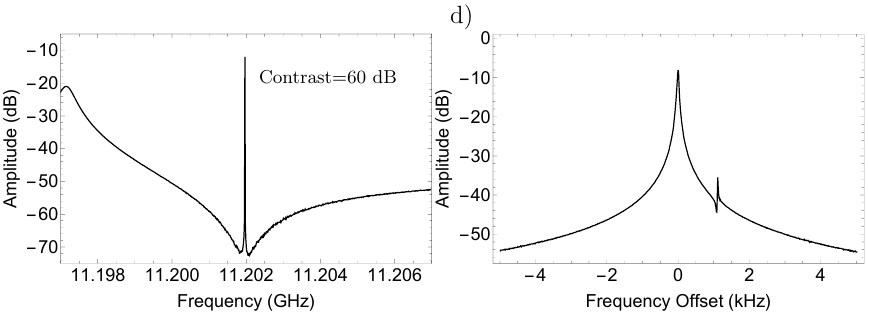}
  \label{Eug6}
\end{minipage}
\caption{Effect of changing the cavity shield dimensions on a high-Q sapphire fundamental quasi-TM WGH$_{16,0,0}$ mode centred at 11.202 GHz. Left, a $10$ $MHz$ span showing the properties of the background mode with contrast. Right, a $10$ kHz span showing the properties of the high-Q mode. a) A cylindrical shielding cavity height of $h\approx 52.10 mm$, b) $\Delta h\approx 60 \mu m$, c) $\Delta h\approx 70 \mu m$, and d) a further $\Delta h\approx 30 \mu m.$ The bottom curve shows unambiguously the best symmetry and contrast occurs in a background anti-resonance. Calculating the Q-factors from Fano fits to the transmission curve leads to a maximum loaded Q-factor of $2.5\times 10^8$ in anti-resonance, however the value only deviated below this by no more than 20 percent across the anti-resonance tuning range.}
\label{ELots}
\end{figure}

\section{Quantifying Symmetry and Contrast}

Due to the importance of high contrast and symmetry for optimising phase-noise performance, in this section the exact process of quantifying these values is described. The contrast values presented in this section, and throughout this paper are calculated by fitting a parabola to the background of any narrow span spectra acquired, where resonant modes or anti-resonance features are removed from the dataset. The maximum dB point of the experimental mode then has the corresponding value of the background at the same frequency subtracted off to provide an estimated value of contrast.

The degree of symmetry of a mode, $\Sigma$, is determined from the transmission function ($S_{12}$), given by
\begin{equation}
\Sigma=\frac{\int_{-10\kappa_{0}}^{0}|S_{12}(f_{0}+f)-\beta_{0}+S_{12}(f_{0}-f)-\beta_{0}|df}{\int_{-10\kappa_{0}}^{0}|S_{12}(f_{0}+f)-\beta_{0}|+|S_{12}(f_{0}-f)-\beta_{0}|df}
\label{differenceeq}
\end{equation}
with $\beta_{0}$ the background term given by \begin{equation}
\beta_{0}=\frac{1}{2}(S_{12}(f_{0}-10\kappa_{0})+S_{12}(f_{0}+10\kappa_{0}))
\end{equation} where $\kappa_{0}$ is the bandwidth of the mode (full-width half-max). Here, $f_{0}$ is the frequency of the local maximum (or minimum for inversions). The integration region of ten bandwidths (10$\kappa_{0}$) is determined to sufficiently represent the mode structure. Since the VNA used in the measurements of $S_{12}(f)$ is digital with a set resolution bandwidth, the data was interpolated and then integrated. The value of $\Sigma$, provides a degree of mode symmetry where an ideal Lorentzian (with Fano parameter $q=0$) has a value of $\Sigma=1$ corresponding to perfect symmetry, while perfect asymmetry will have a value of  $\Sigma=0$. 

\section{Optimizing Contrast and Symmetry of a Cryogenic Sapphire  Loaded Cavity}

We investigate background anti-resonance tuning by changing the shield dimensions of a high-Q sapphire WG mode resonator as shown in Fig.~\ref{EugCavity}. High quality-factor modes exist amongst background resonances and anti-resonances. The key feature in this case is that the background anti-resonances and resonances are tuned by varying the cavity height, $h$ (dimension as shown in Fig.~\ref{ELots}), whilst the high-Q resonances depend only on the crystal dimensions, as they are overwhelmingly located within the dielectric.


With each sequential change in the cavity height a frequency shift in the relatively broad background modes was observed, on the order of 30 kHz per $\mu$m for the change in cavity height. In this case, it was therefore possible to shift an anti-resonance directly onto the observed mode. As the background anti-resonance was tuned over the high-Q mode, distortion of the high-Q mode was observed, with a large change in both contrast and symmetry. This affect may be explained by the onset of a Fano-resonance, as seen in Fig.~\ref{ELots}, modelled through a mutual resistive term between the two modes as described in \cite{tobar1991generalized}. This result was reproduced with a second sapphire resonator at a similar anti-resonance position, with the sapphire's orientation chosen to maximise probe coupling to one of the doublets and minimise it to the other. The small finite coupling to the second under coupled mode causes some slight variations to the effect. These two resonators are being developed to create high-stability and low noise cryogenic sapphire oscillators \cite{ivanov2021noise,10286875}, so it is important to gain good symmetry and contrast in order to achieve optimum phase noise performance.

\section{Wire Dipole Probe}

\begin{figure}[t]
  \centering
  \includegraphics[width=1.0\linewidth]{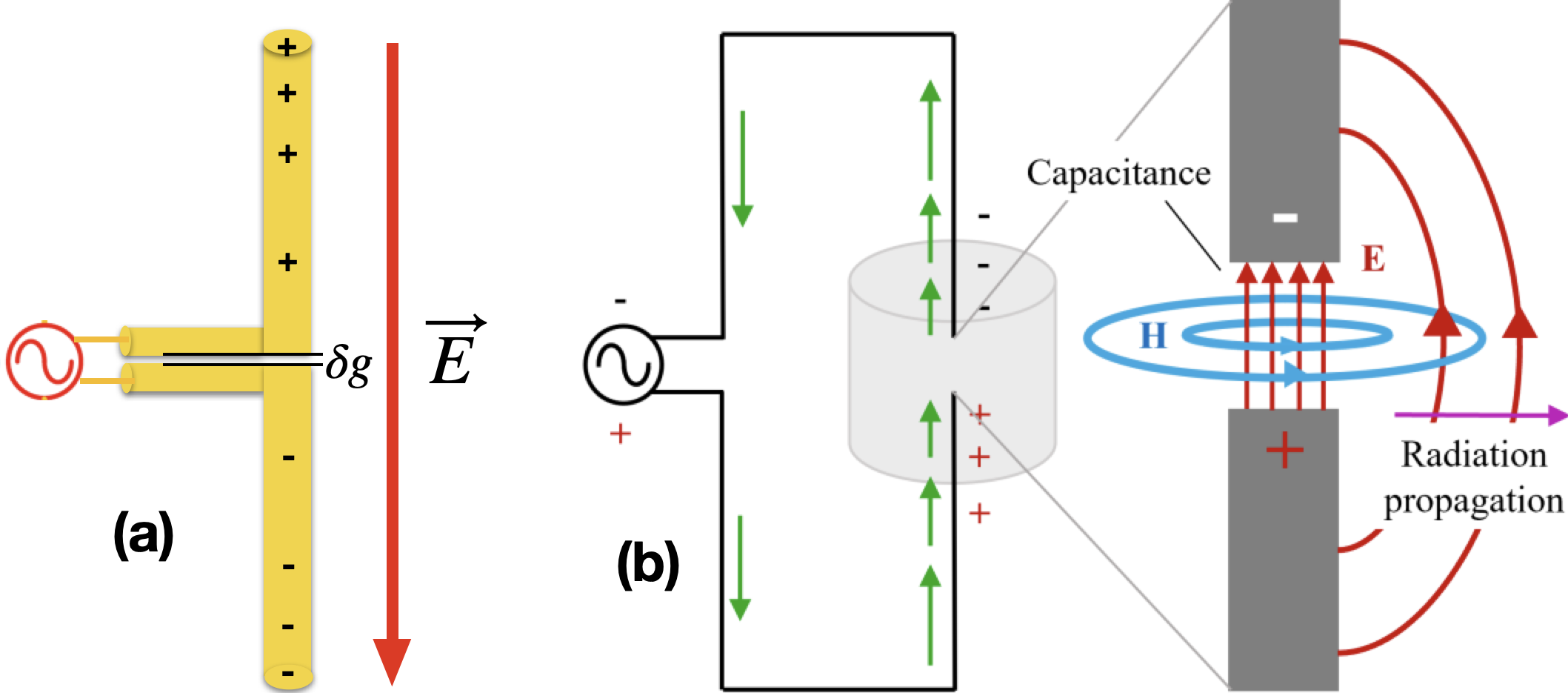}
  \caption{a) Dipole antenna driven by a voltage source through a sub-wavelength gap, so the current and voltage across the antenna is out of phase, and hence capacitive. Likewise the near electric $(\vec{E})$ and magnetic field $(\vec{H})$ are out of phase. b) The modified dipole antenna arrangement for use as a coupling probe to a cavity resonator (light grey cylinder). Right is a magnification of the gap between the antenna arms illustrating the capacitive coupling.}
  \label{DipoleTypes}
\end{figure}

To measure or excite a normal electric field, an electric field probe may be constructed from a 50 $\Omega$ open-end coaxial cable oriented in a parallel direction to this field, with the central conductor extruded directly out of the cable.  Constructing an antenna that measures or excites the tangential field is more complicated as the tangential field in principle goes to zero for a perfect conducting probe. However, this may be achieved by using a balanced wire dipole probe, made with two non-contacting adjacent electric field probes, as shown in Fig.~\ref{DipoleTypes}a). The dipole antenna nominally consists of two wires with a sub-wavelength gap $\delta$g, through which the electric field is excited or measured. The gap is much smaller than the antenna dimensions, and provides capacitive coupling. 

To adapt this type of antenna to a cylindrical cavity resonator the probe may be coupled through the top and bottom of the cylinder as shown in Fig.~\ref{DipoleTypes}b), combined with a hybrid 180-degree junction or balun to balance the power and phase to the two arms of the dipole probe antenna \cite{laurin2001near,baudry2007applications,yang2022dual,tice1955probes,s22187029}. The phase and attenuation between the two adjacent coaxial cables of the dipole wire probe give extra degrees of freedom that may be varied while coupling to the resonator's modes.

\begin{figure}[t]
  \centering
  \includegraphics[width=1.0\linewidth]{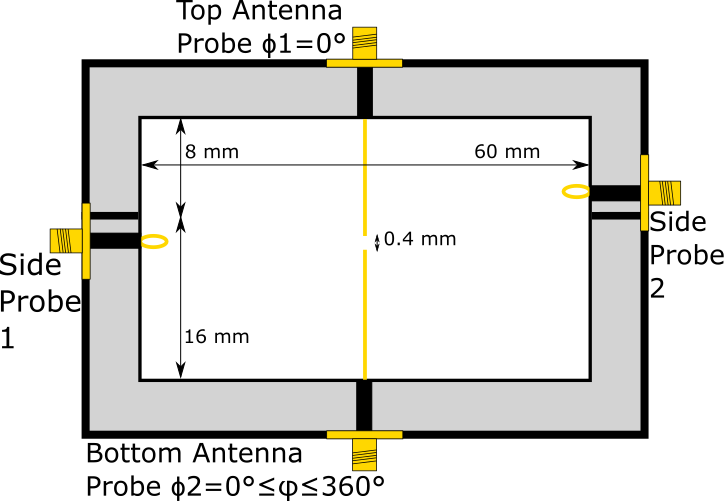}
\caption{The empty copper cylindrical cavity with dimensions and probe positions, with central conductors and SMA ports indicated in yellow.}
\label{empty}
\end{figure}

\begin{figure}[h]
\centering
\begin{minipage}{.4\textwidth}
\includegraphics[width=1\linewidth]{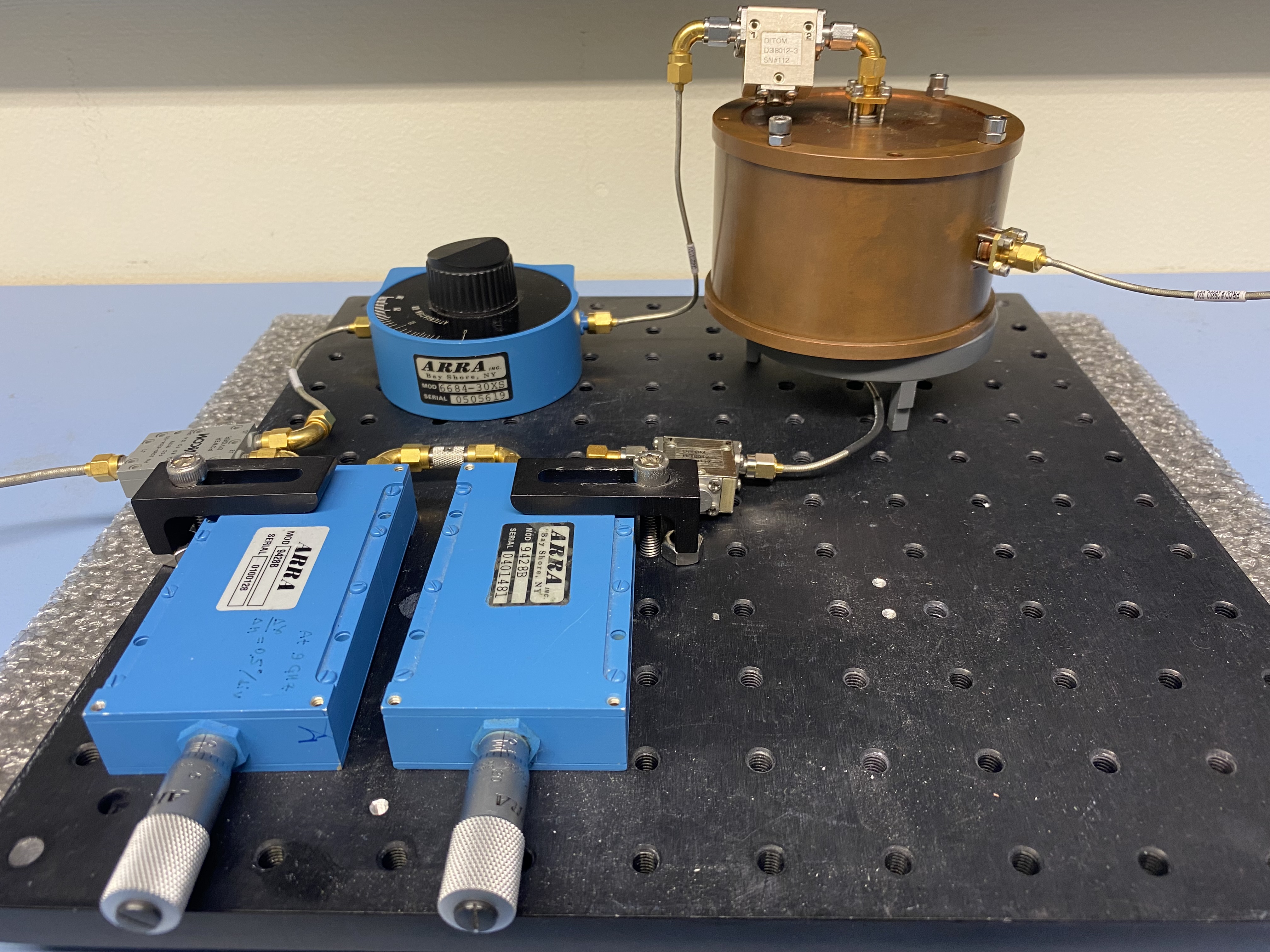}
\end{minipage}
\begin{minipage}{.4\textwidth}
\includegraphics[width=1\linewidth,angle=0]{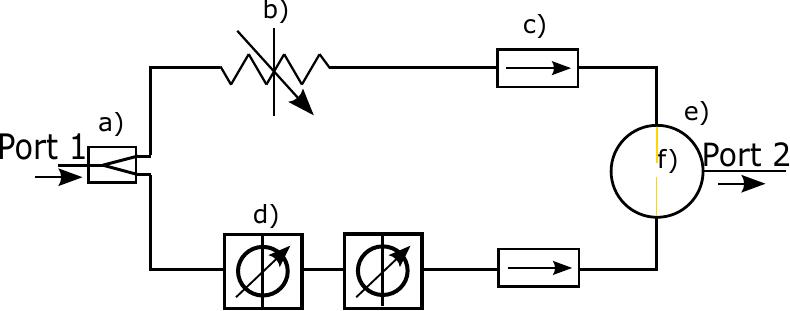}
\end{minipage}%
\caption{Top, photograph of the microwave interferometeric circuit. Bottom, a schematic illustrating the components: a) power splitter, b) variable attenuator, c) isolator, d) phase shifter, e) cavity resonator and f) dipole wire probes. The setup is similar to a Mach Zehnder interferometer, except that it measures resonance and anti-resonance instead of bright and dark ports. For network analyser experiments port 1 is labelled as the input of the interferometer, whilst probe 2 is labelled as the output probe on the side of the cavity.}
\label{Setup}
\end{figure}

\begin{figure}[t]
\begin{minipage}{.25\textwidth}
  \centering
  \includegraphics[width=1\linewidth]{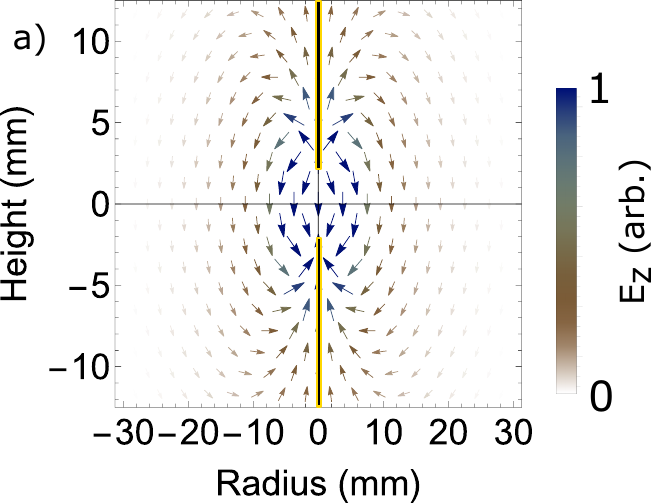}
  \label{gooddipolefield}
\end{minipage}%
\begin{minipage}{.25\textwidth}
  \centering
  \includegraphics[width=1\linewidth]{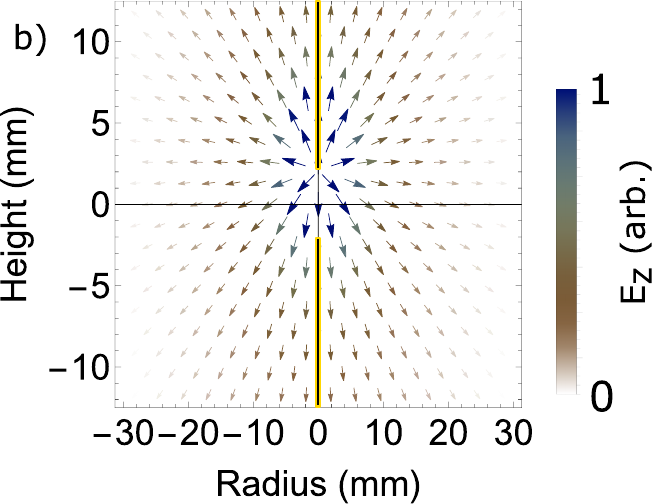}
  \label{hybriddipolefield}
\end{minipage}
\centering
\begin{minipage}{.25\textwidth}
  \centering
  \includegraphics[width=1\linewidth]{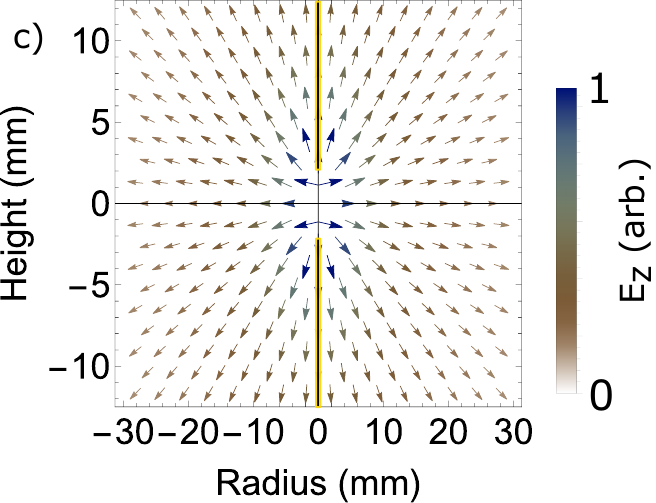}
  \label{goodmonofield}
\end{minipage}
\caption{The wire dipole antenna can be configured to produce various electric field patterns depending on the relative phase difference $\Delta \phi$ between the two arms. The three field configurations for a fixed time, $t$, are a) dipole with $\Delta \phi=\pi$, b) hybrid with $\Delta \phi=\frac{\pi}{2}$, and c) monopole with $\Delta \phi=0$, with probe positions drawn in yellow.}
\label{dipoletypes}
\end{figure}

\section{Microwave dipole probe interferometer}
To use such a wire dipole antenna to excite an empty cylindrical cavity resonator, it was set up symmetrically about the centre of the cavity, as shown in Fig.~\ref{empty}, with so-called top and bottom  as depicted. Control over the phase and attenuation difference between the top and bottom antennas can be achieved by use of elements similar to a Mach Zehnder interferometer \cite{born2013principles} as shown in Fig.~\ref{Setup}.

The antenna can produce dipole, monopole or hybrid field patterns as depicted in Fig.~\ref{dipoletypes}, determined by the phase difference $\Delta \phi=\phi_{2}-\phi_{1}$ between the top and bottom probe. For a dipole configuration we set to $\Delta \phi=\pi$. The depicted setup in Fig.~\ref{Setup} is in fact more general and allows  $\Delta \phi$ to take any value between $0$ and $2\pi$. Two phase shifters combined in series after a 3 dB power splitter are necessary to ensure at least $2\pi$ phase shift in the frequency range of interest, and utilised micrometers to vary the phase shift (line stretchers). Henceforth, relative phase variations in this work are stated in terms of mm. Due to the slight attenuation of these phase shifters, small variable and static attenuators were used to balance out the attenuation in the arms of the interferometer. It is critical that both input probes of the cavity were at precisely the same power to ensure effectiveness of the dipole antenna as a function of $\Delta \phi$. Isolators are employed in each arm of the interferometer to function as a voltage wave standing ratio (VSWR) elimination mechanism. VSWR causes reflections in the transmission lines and accentuates features of the circuit setup, rather than the dipole antenna. Thus it was desirable that this VSWR was eliminated.

As illustrated in Fig.~\ref{dipoletypes}, the introduction of a phase shift between the two dipole arms affects the orientation and magnitude  of the electric field that the probe couples to. This affects the coupling of the dipole probe to various cavity modes by simply changing the interferometer parameters, and is dependent on the mode's polarisation, as well as the extent of the coupling to cross-talk  \cite{PhysRevApplied.13.044039,PhysRevApplied.19.014030} between the dipole probe and the output side probe. Anti-resonances will form due to destructive interference between resonant modes and either the input driving field or the cross-talk, therefore tuning the phase shift across the dipole probe can effectively change the spectral location and extent of destructive and constructive interference \cite{Rao_2019}. Thus, anti-resonance and resonance frequencies may be tuned via the phase difference imposed by the interferometer. So in contrast to the first method of tuning anti-resonances by cavity height, this novel cavity excitation method allows background anti-resonances to be tuned by the non-destructive method of varying the phase difference between the input probe arms.

\subsection{Operation and Calibration of a Cylindrical Cavity Resonator}

\begin{figure}[t!]
  \centering
  \includegraphics[width=1\linewidth]{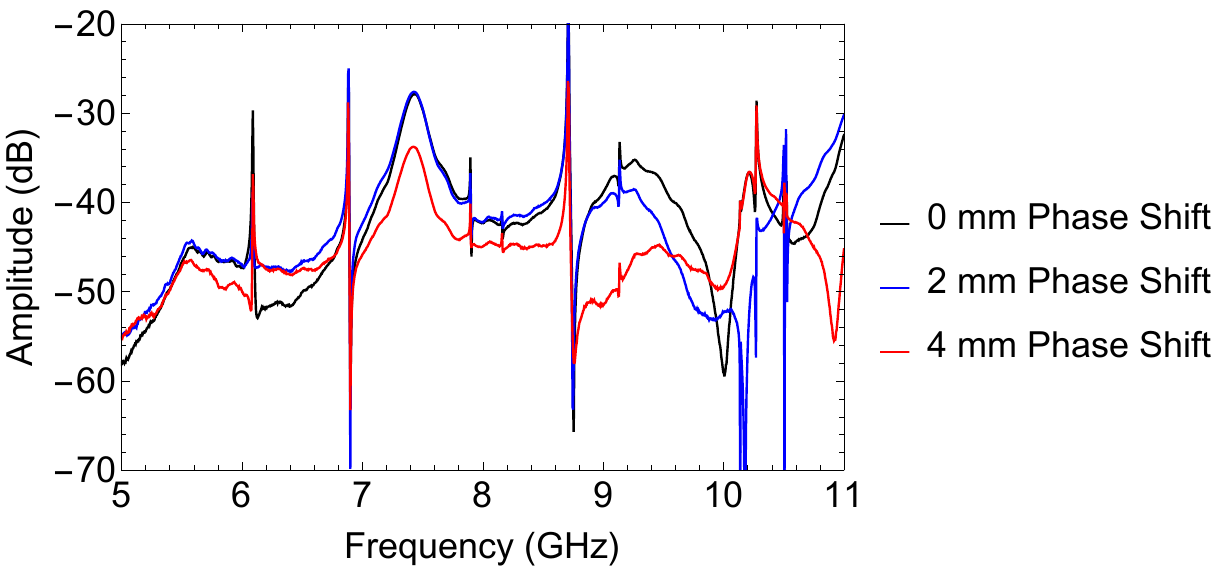}
  \includegraphics[width=1\linewidth]{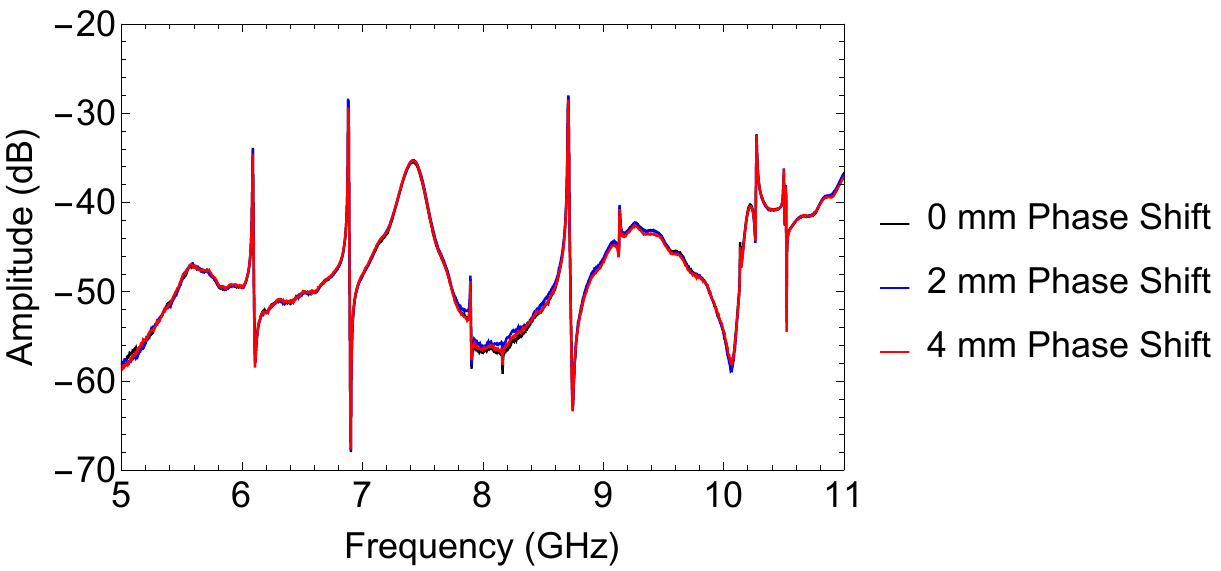}
  \caption{The 8 GHz centre transmission readout at three arbitrary, equally spaced phase shifts with a span of 6 GHz.  Top: The spectrum when the power in both arms of the interferometer driving the dipole antenna have no attenuation added (0 dB). Bottom: The spectrum when one arm of the interferometer driving the dipole antenna was attenuated by 30 dB, where the attenuation is controlled using the variable attenuator in Fig.~\ref{Setup}. When both arms of the interferometer are activated, the appearance of broad and sharp anti-resonances occur and can bee seen in the 9-10 GHz and 10-11 GHz regions. When one arm is deactivated with 30 dB attenuation, the phase shift of the remaining interferometer arm makes no difference to the recorded spectrum, indicating that there is no more interference.}
  \label{7.5G0and30Attenuation}
\end{figure}

The configuration of the empty cylindrical cavity resonator examined in this section is shown in Fig.~\ref{empty}. Unlike the sapphire-loaded cavity of Fig.~\ref{EugCavity}, it is impossible to change the dimensions of a cavity resonator to tune the background modes without tuning the the high-Q resonant mode itself, as it too is defined by the cavity dimensions.

Here, we examine the empty copper cylindrical cavity resonator to gain an understanding of the transmission background coupled to the dipole interferometric antenna. Transmission spectra from 5-11 GHz are examined as the phase difference is varied, as shown in Fig.~\ref{7.5G0and30Attenuation}.

\begin{figure}[t!]
  \centering
  \includegraphics[width=0.9\linewidth]{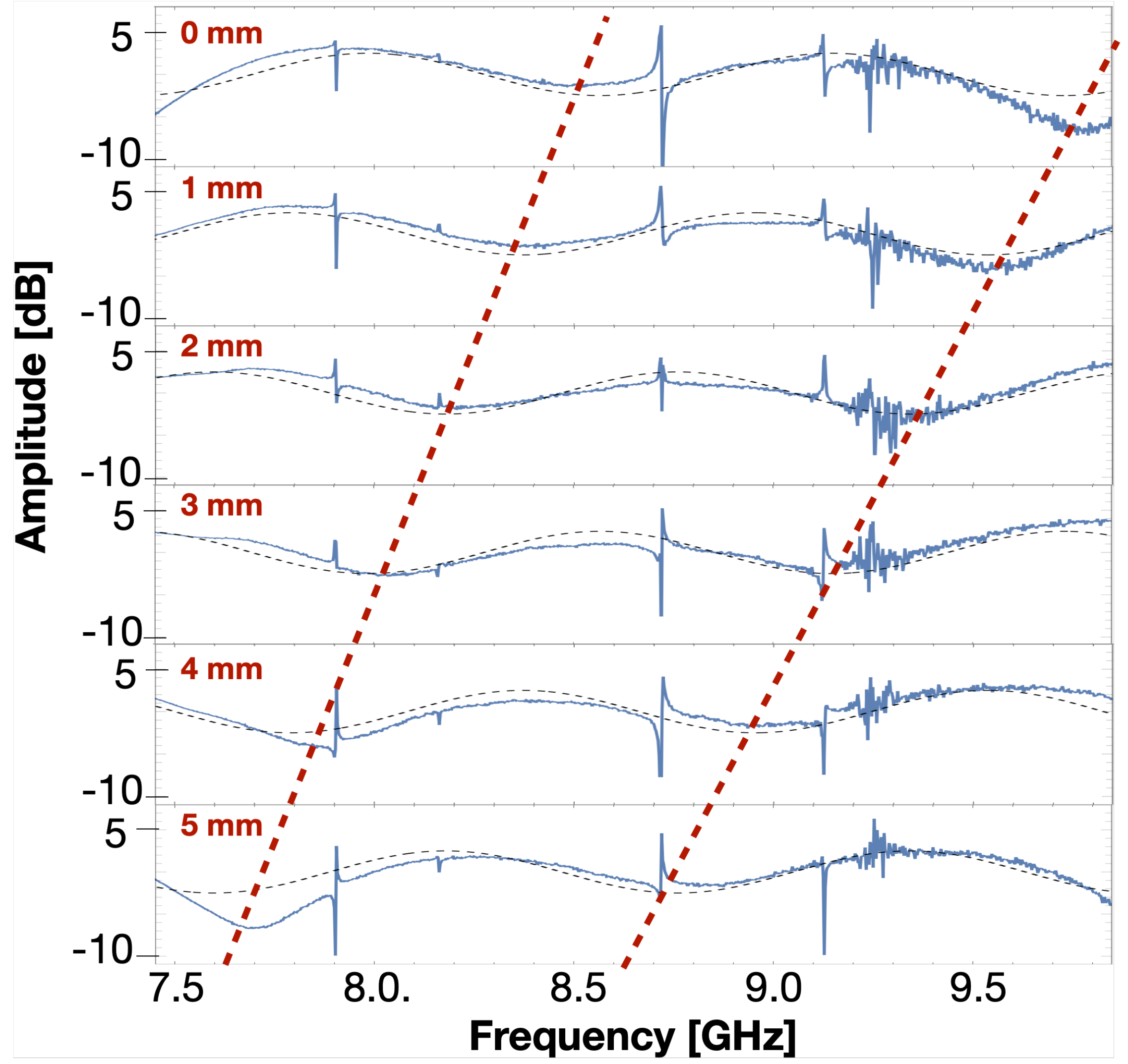}
\caption{Calibration graph for frequency-dependent phase shift measurements. Six plots of $S_{12}$ measurement data for each phase subtracted from the averaged transmission data across all six phases (0 mm to 5 mm in descending order). The black-dashed line indicates a simple sinusoidal fit to help indicate the tuning of an anti-resonance. The red dashed lines indicate the actual anti-resonance tuning. These curves can be used to calibrate the frequency phase shift of the back ground modes. For example, at 8 GHz it roughly takes 4 mm of turning to go from resonance to anti-resonance ($\sim\pi$ phase shift).}
\label{machp0}
\end{figure}

Two primary characteristics of the spectrum are, 1) broad low-Q background resonances/anti-resonances, and 2) the appearance of sharp anti-resonances. To deduce that the tuning of background resonances and anti-resonances were in fact related to the phase shift inside the dipole probe interferometer, one antenna arm was deactivated by severe attenuation, whilst the other was phase shifted, and compared when the attenuation was set to zero, with the results plotted in Fig.~\ref{7.5G0and30Attenuation}. Deactivating one arm produced no phase-dependent effects, with three identical transmission spectra at different values of phase showing virtually no difference. On the contrary, with two active arms (attenuation set to zero), the dipole antenna produced significant changes in the spectra, particularly in the 9-10 GHz region. This result suggests that frequency shifting of background resonances/anti-resonances were caused by varying the phase difference ($\Delta \phi$) between the two arms of the active dipole antenna.

The use of the dipole probe in this manner allows for tuning of the background mode \textit{without} frequency tuning the higher Q mode. This allows similar control over the experiment without physically changing the cavity dimensions as in the prior case. Fig.~\ref{machp0}, shows the distinct interferometric pattern that gives rise to the tuning of the back ground resonances and anti-resonances. This figure can also be used to calibration the sensitivity of the phase shifters. One can see that the operational modes have their best transmission properties when in a background anti-resonance. In the next sections we show they have a better Q-factor and contrast.

\section{Dipole Interferometric Antenna: Cavity Mode Excitation}

\subsection{Tuning a Low-Q Anti-Resonance}

\begin{figure}[h]
\begin{minipage}{.5\textwidth}
  \centering
  \includegraphics[width=1\linewidth]{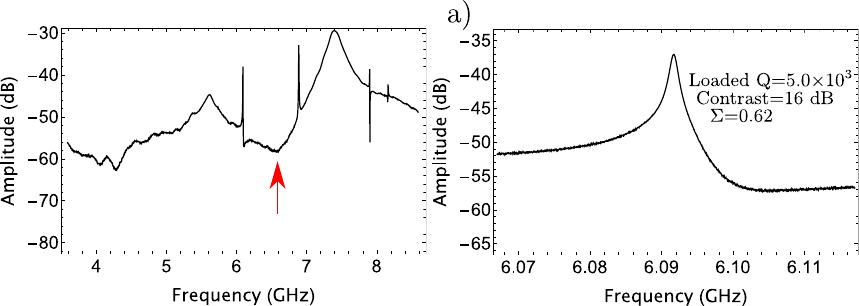}
  \label{609righttwo}
\end{minipage}

\begin{minipage}{.5\textwidth}
  \centering
  \includegraphics[width=1\linewidth]{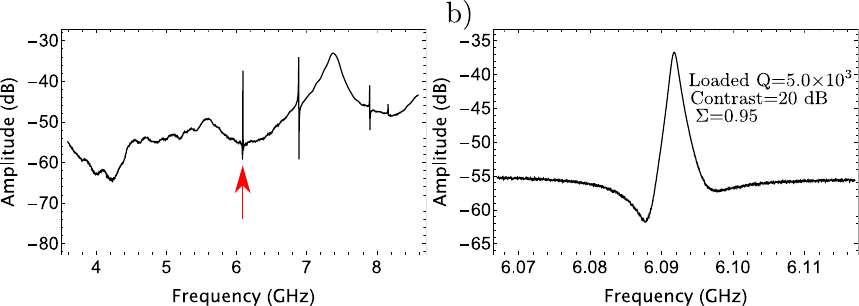}
  \label{609centredtwo}
\end{minipage}
\begin{minipage}{.5\textwidth}
  \centering
  \includegraphics[width=1\linewidth]{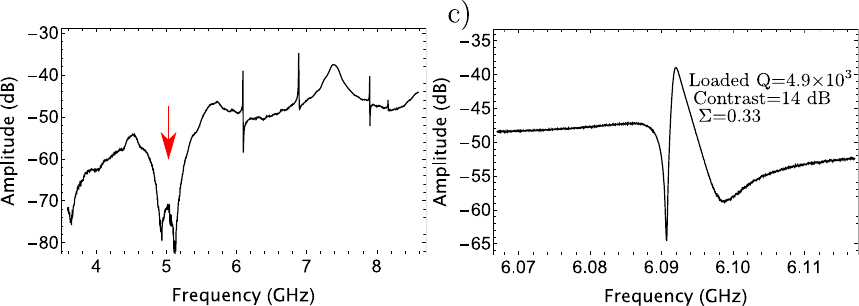}
  \label{609opptwo}
\end{minipage}
\caption{A broad 5 GHz span (left) and narrow 50 MHz span (right) of the $TM_{1,1,0}$ 6.09 GHz mode transmission spectra at varying phase shift and constant attenuation. The mode is displayed when the anti-resonance is positioned: a) to the right of mode, b) centrally on the mode, and c) to the left of the mode. The anti-resonance positions are indicated by a red arrow. The loaded Q, contrast and symmetry values are given for the respective anti-resonance positions.}
\label{609}
\end{figure}

Fig.~\ref{609}, shows the tuning of a broad anti-resonance by varying ($\Delta \phi$), with the effect of the anti-resonance's in proximity to the $TM_{1,1,0}$ 6.09 GHz mode shown at four unique positions.

The maximum mode symmetry in Fig.~\ref{609} (on anti-resonance) was achieved by tuning the broad anti-resonance directly on to the mode of interest, done so by varying the phase of one of the antenna arms. There is also a significant increase in contrast of the mode with the background when this occurs, changing from a minimum contrast of 13 dB (anti-resonance tuned away), to 20 dB (anti-resonance tuned on to the mode). Tuning the anti-resonance on to the mode of interest produced a higher degree of symmetry. The broad span in Fig.~\ref{609} on anti-resonance helps to visualise this on a larger scale, where the mode of interest is now observed to be in the well of an anti-resonance, with a lower transmission background.






\subsection{Tuning a Higher-Q Anti-Resonance}
\begin{figure}[ht!]
  \centering
  \includegraphics[width=1\linewidth]{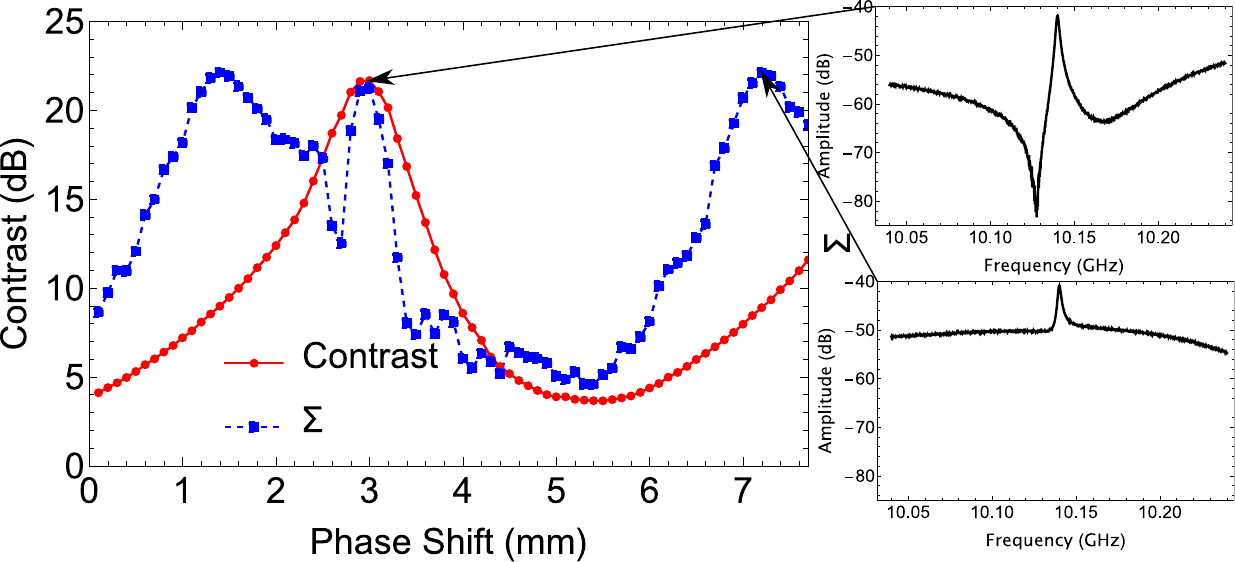}
  \label{10consym}
\caption{The optimisation of the $TM_{2,1,1}$} 10.14 GHz mode when centred on antiresonance. The contrast (red, solid) and degree of symmetry, $\Sigma$ (blue, squares) is plotted using a double y-axis. Peak symmetry and contrast coincided near 3 mm phase shift. The two local maximum transmission spectra are shown with their corresponding locations. The spectra positioned in anti-resonance has $\approx$3 times better contrast. The Q-factor of the mode is independent on phase shift and equal to 4,600.
\label{10Character}
\end{figure}

Among the broad tuning background anti-resonances, there is also the presence of a few sharp or deeper anti-resonances, similar to that observed in the cryogenic sapphire resonator in the prior section. It was possible to tune one of these sharp anti-resonances to the same frequency of the $TM_{2,1,1}$ 10.14 GHz mode. This led to an optimisation of contrast and symmetry shown in Fig.~\ref{10Character}.

The contrast and symmetry are both maximised when the sharp anti-resonance is positioned directly on the resonant mode centre near 3 mm phase shift in this case, noting that the peak here is possibly underestimated due to the limited data resolution around this applied phase shift of the interferometer. The results show an increase of over 18 dB in contrast at the second point of maximum symmetry. Other high symmetry peaks are also present in Fig.~\ref{10Character} as expected, which occurs when the background resonance  is tuned to the $TM_{2,1,1}$ 10.14 GHz mode, but in such situations the contrast is poor due to the interference with the background resonance.

\section{Dipole Interferometric Antenna: Sapphire Whispering Gallery Mode Excitation}

\begin{figure}[h!]
\centering
\begin{minipage}{.30\textwidth}
\includegraphics[width=1\linewidth]{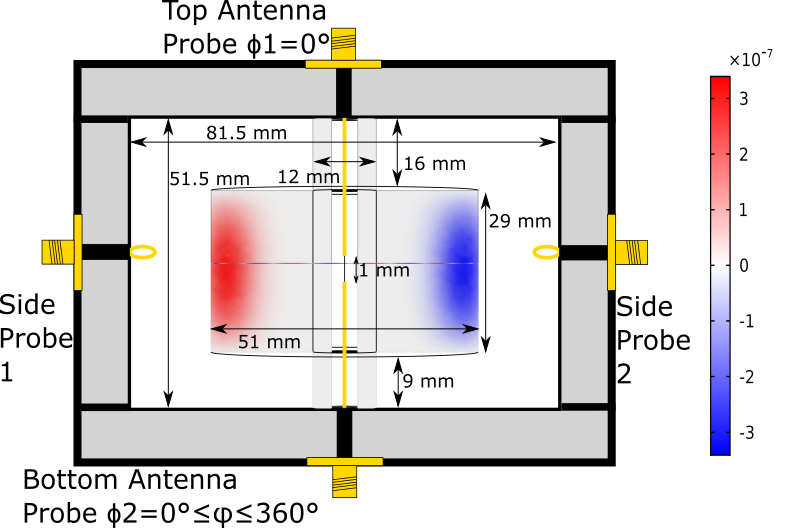}
\end{minipage}%
\begin{minipage}{.24\textwidth}
\includegraphics[width=1\linewidth,angle=-90]{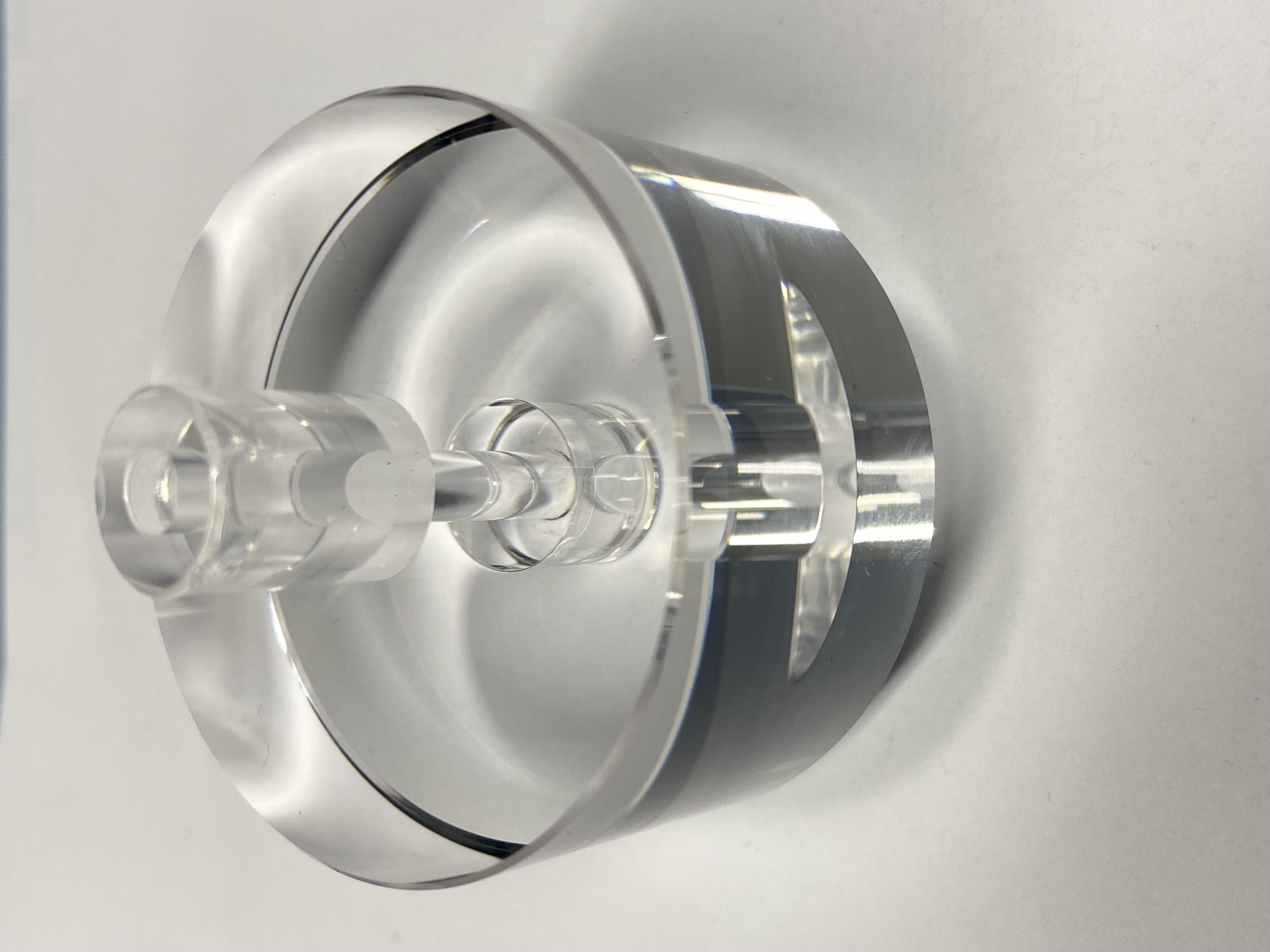}
\end{minipage}
\caption{Left, the copper cavity and sapphire placed inside with associated $D_{z}$ field cross-section for the m=11 whispering gallery mode. Right, a photo of the sapphire with cylindrical spindle. The gap between the dipole antenna probes was kept around 1 mm for the sapphire cavity.}
\label{cavsaph}
\end{figure}

In this section we show room temperature results of when the dipole antenna was coupled to a high-Q sapphire whispering gallery cavity as shown in Fig.~\ref{cavsaph}, with similar supporting interferometric circuitry. These modes are more than an order of magnitude higher-Q than the empty copper cavity modes. In this higher Q system, several modes were examined, which displayed interesting behaviour when tuning the phase of the interferometer.

\subsection{Tuning Whispering Gallery Mode Type 1}
\begin{figure}[t!]
\begin{minipage}{.25\textwidth}
  \centering
  \includegraphics[width=1\linewidth]{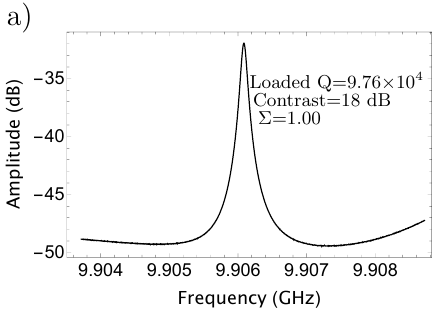}
  \label{Saph9.905On}
\end{minipage}%
\begin{minipage}{.25\textwidth}
  \centering
  \includegraphics[width=1\linewidth]{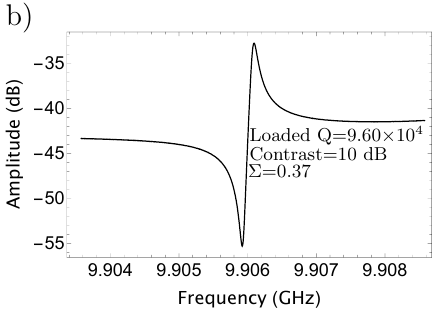}
  \label{Saph9.905Right}
\end{minipage}
\begin{minipage}{.25\textwidth}
  \centering
  \includegraphics[width=1\linewidth]{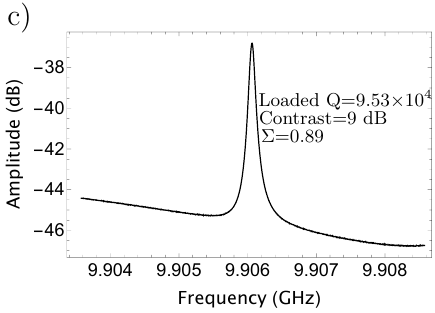}
  \label{Saph9.905Off}
\end{minipage}%
\begin{minipage}{.245\textwidth}
  \centering
  \includegraphics[width=1\linewidth]{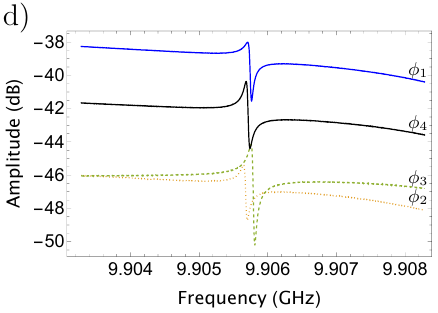}
  \label{SaphLowCouple}
\end{minipage}
\caption{The 9.905 GHz sapphire mode interacting with a background mode, with a narrow span a) centred on the anti-resonance, b) with the anti-resonance to the left of mode, and c) centred on the background resonance. Spectra a-c have the side probe coupling set to near 0.94, although the anti-resonance centred spectrum is slightly less coupled at 0.81. d) Four transmission spectra in a low-coupled system with side-probe coupling near 0.1. Broadband transmission changes are observed for phase positions $\phi_{1-4}$, and the ability to tune broadband anti-resonance is suppressed. The loaded Q, contrast, and symmetry values ($\Sigma$) are presented for each anti-resonance position.}
\label{Saph9.905}
\end{figure}

The first whispering gallery (WG) mode analysed, was centred at 9.905 GHz, as shown in Fig.~\ref{Saph9.905}. From COMSOL finite element modelling, we identify this mode as WGH$_{11,1,0}$. The results also show that the optimal symmetry and contrast occured when the operational mode was tuned within a background anti-resonance. The sapphire mode's contrast improved by several dB, along with the line shape symmetry when positioned on an anti-resonance. Along with these results a Q-factor increase of about $2\%$ was observed. The effect of the side-probe coupling on the dipole tuning features of this mode were also examined in Fig.~\ref{Saph9.905}.

The coupling data from Fig.~\ref{Saph9.905} indicate a less pronounced anti-resonance effects at lower output couplings when the contrast is low. When the probe is retracted, it is less coupled to the high-Q sapphire modes and detects less of this influence to background resonances. In contrast, when the output probe is near-unity coupled the effects are more pronounced. To achieve good phase noise in an oscillator high contrast is necessary and one usually designs the probes to be near-unity coupled.

\subsection{Tuning Whispering Gallery Mode Type 2}
\begin{figure}[t!]
\begin{minipage}{.25\textwidth}
  \centering
  \includegraphics[width=1\linewidth]{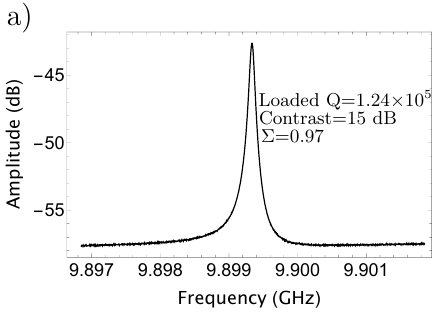}
  \label{Saph9.899CloseOn}
\end{minipage}%
\begin{minipage}{.25\textwidth}
  \centering
  \includegraphics[width=1\linewidth]{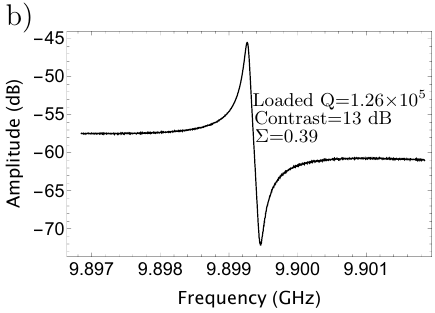}
  \label{Saph9.899CloseRight}
\end{minipage}
\begin{minipage}{.25\textwidth}
  \centering
  \includegraphics[width=1\linewidth]{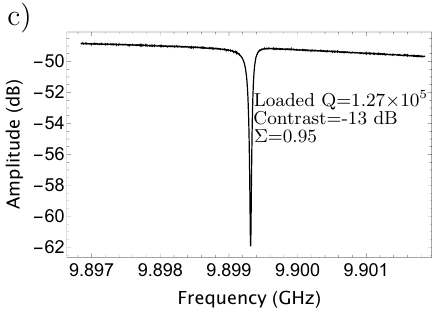}
  \label{Saph9.899CloseOff}
\end{minipage}%
\begin{minipage}{.25\textwidth}
  \centering
  \includegraphics[width=1\linewidth]{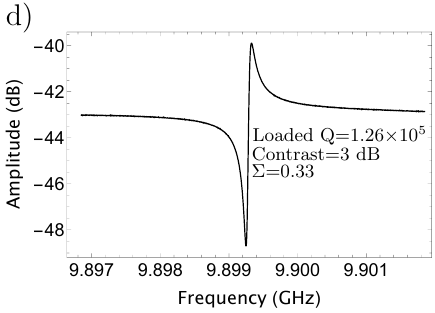}
  \label{Saph9.899CloseLeft}
\end{minipage}
\caption{The 9.899 GHz sapphire mode interacting with a background mode, a) centred on the anti-resonance, b) with the anti-resonance to the right of mode, c) centred on the background resonance, and d) with the anti-resonance to the left of sapphire mode (bottom right). The loaded Q, contrast, and symmetry values ($\Sigma$) are presented for each anti-resonance position.}
\label{Saph9.889Small}
\end{figure}

The properties of tuning an anti-resonance to second high-Q sapphire mode is shown in Fig.~\ref{Saph9.889Small}. From finite element modelling we identify this mode as a fundamental quasi TM WG mode, typically denoted as WGH$_{14,0,0}$ with 96 percent of the electric field energy in the axial z-direction in sapphire and thus it strongly couples to the dipole probe, which is oriented in the axial z-direction. The mode displays features consistent with the dipole anti-resonance tuning, with the additional feature of a complete mode inversion when the sapphire mode is positioned on resonance with respect to the background mode, similar to electromagnetically induced transparency \cite{tobar1991generalized}. This inversion point is also a point of maximum symmetry however has reduced magnitude of contrast.

%

A third mode investigated is a fundamental quasi TM WG mode, typically denoted as WGH$_{11,0,0}$ with 93 percent of the electric field energy in the axial z-direction in sapphire, and displayed similar properties to the previous one.

\section{Conclusion}
In this paper, two techniques of tuning anti-resonances within a microwave cavity system were investigated, including a novel Mach-Zehnder configuration to excite microwave cavities with a generalized dipole antenna configuration. The centring of coupled background anti-resonances onto the modes of interest produced large improvements in mode symmetry and contrast. These features are desirable in many applications such as oscillator phase noise performance.


\section*{Acknowledgments}
This work was funded by the Australian Research Council Centre of Excellence for Engineered Quantum Systems, CE170100009 and  the Australian Research Council Centre of Excellence for Dark Matter Particle Physics, CE200100008. Thank you to Misty Lakelin for contributing the dipole antenna diagram in Fig.~\ref{DipoleTypes}.



\end{document}